\documentclass[useAMS,usenatbib]{mn2e}
\usepackage{natbib}
\usepackage{epsf}
\usepackage{graphicx}

\title[An optical--X-ray correlation in 4U 1957+11]{A long-term optical--X-ray correlation in 4U 1957+11}
\author[D. M. Russell et al.]{D. M. Russell$^{1}$\thanks{E-mail: d.m.russell@uva.nl}, F. Lewis$^{2,3,4}$, P. Roche$^{2,3,4}$, J. S. Clark$^{3}$, E. Breedt$^{5}$, R. P. Fender$^{5}$\\
$^{1}$Astronomical Institute `Anton Pannekoek', University of Amsterdam, P.O. Box 94249, 1090 GE Amsterdam, the Netherlands\\
$^{2}$Faulkes Telescope Project, School of Physics and Astronomy, Cardiff University, 5 The Parade, Cardiff, CF24 3AA, UK\\
$^{3}$Department of Physics and Astronomy, The Open University, Walton Hall, Milton Keynes, MK7 6AA, UK\\
$^{4}$Division of Earth, Space \& Environment, Univ. of Glamorgan, Pontypridd, CF37 1DL, UK\\
$^{5}$School of Physics \& Astronomy, University of Southampton, Highfield, Southampton, SO17 1BJ, UK\\
}
\begin{document}


\pagerange{\pageref{firstpage}--\pageref{lastpage}} \pubyear{2008}

\maketitle

\label{firstpage}

\begin{abstract}
Three years of optical monitoring of the low-mass X-ray binary (LMXB) 4U 1957+11 is presented. The source was observed in $V$, $R$ and $i$-bands using the Faulkes Telescopes North and South. The light curve is dominated by long-term variations which are correlated (at the $> 3 \sigma$ level) with the soft X-ray flux from the All-Sky Monitor on board the Rossi X-ray Timing Explorer. The variations span one magnitude in all three filters. We find no evidence for periodicities in our light curves, contrary to a previous short-timescale optical study in which the flux varied on a 9.3-hour sinusoidal period by a smaller amplitude. The optical spectral energy distribution is blue and typical of LMXBs in outburst, as is the power law index of the correlation $\beta = 0.5$, where $F_{\rm \nu,OPT}\propto F_{\rm X}^{\beta}$. The discrete cross-correlation function reveals a peak at an X-ray lag of 2--14 days, which could be the viscous timescale. However, adopting the least squares method we find the strongest correlation at a lag of $0 \pm 4$ days, consistent with X-ray reprocessing on the surface of the disc. We therefore constrain the optical lag behind X-ray to be between -14 and +4 days. In addition, we use the optical--X-ray luminosity diagram for LMXBs as a diagnostic tool to constrain the nature of the compact object in 4U 1957+11, since black hole and neutron star sources reside in different regions of this diagram. It is found that if the system contains a black hole (as is the currently favoured hypothesis), its distance must exceed $\sim 20$ kpc for the optical and X-ray luminosities to be consistent with other soft state black hole systems. For distances $< 20$ kpc, the data lie in a region of the diagram populated only by neutron star sources (black hole systems are ten times optically brighter for this X-ray luminosity). 4U 1957+11 is unique: it is either the only black hole LMXB to exist in an apparent persistent soft state, or it is a neutron star LMXB which behaves like a black hole.
\end{abstract}


\section{Introduction}

4U 1957+11 (V1408 Aql) is a persistently active ($\sim$ 1--10$\times 10^{-10}$ erg cm$^{-2}$ s$^{-1}$; 1.5--12 keV) low-mass X-ray binary (LMXB). Its X-ray spectrum has always appeared soft since its discovery \citep*{giacet74,wijnet02}. These two properties make this source unique; most LMXBs are transient or regularly perform transitions to harder X-ray states \citep*[for more on X-ray states see e.g.][]{hasiva89,homabe05,mcclet06,fendet09}. The X-ray spectrum of 4U 1957+11 can be described by a blackbody plus a power law, with fluxes which vary on timescales of hours to months \citep*{yaqoet93,singet94,nowawi99}.
An apparent 117 day X-ray period \citep{nowawi99} possibly due to a precessing accretion disc viewed almost edge-on was later shown to be much more complex; the long-term X-ray flux variations are likely to be driven by changes in the mass accretion rate \citep{yaqoet93,wijnet02}.

\begin{figure}
\centering
\includegraphics[width=7cm,angle=0]{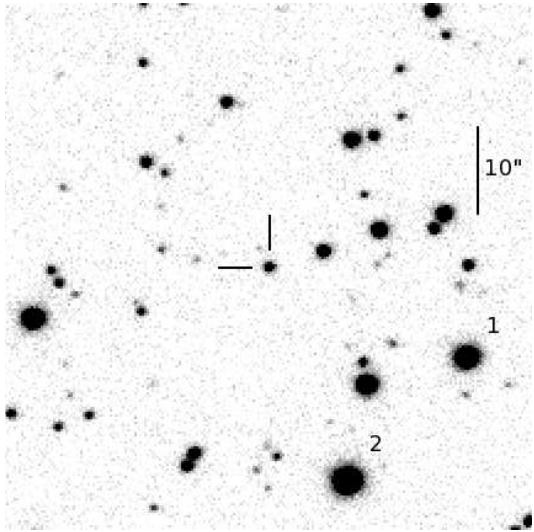}
\caption{Finding chart for 4U 1957+11, indicating the two stars used for flux calibration. The image is a 100 sec exposure in i-band taken with the EM01 camera on FTN under good seeing conditions. The field-of-view is 60 $\times$ 60 arcsec. The optical counterpart of 4U 1957+11 is marked in the centre. The magnitudes of stars 1 and 2 are given in Table 2. North is to the top and east is to the left. For a larger field, see \citeauthor{marget78} (1978).
}
\end{figure}

No estimates of the mass function \citep[e.g.][]{casa07} have been established because the system has never faded to quiescence since its discovery, and type I X-ray bursts typical of neutron stars have not been detected -- as such it is uncertain whether the compact object in the 4U 1957+11 system is a black hole or a neutron star. It was argued \citep{yaqoet93} that the X-ray luminosity is too low, and the inner disc radius too large for a candidate black hole X-ray binary (BHXB), but this may not be the case, and several X-ray spectral and timing properties are similar to other BHXBs \citep{wijnet02,nowaet08}. The X-ray spectrum is typical of soft state BHXBs, and its behaviour shares many similarities with LMC X--3 \citep[][and references therein]{wijnet02}, which contains a $> 5.8$ M$_{\odot}$ black hole and is also in a persistent soft state \citep[but unlike 4U 1957+11, LMC X--3 occasionally performs transitions to the hard state;][]{wilmet01}. Moreover, the luminosity is higher and consistent with soft state BHXBs if the distance to the source is further than the often assumed 7 kpc \citep*[e.g.][]{marget78}. The X-ray spectrum also appears slightly harder at higher fluxes, which is not usually the case for neutron star X-ray binaries \citep[NSXBs;][]{wijnet02}.

Recently, \cite{nowaet08} fitted high-resolution X-ray spectra of 4U 1957+11 with an inclined disc blackbody with an interstellar absorption of n$_{\rm H} = (1.5 \pm 0.5) \times 10^{21}$ cm$^{-2}$. The total absorption through the Galaxy in the direction of the source is $1.2 \times 10^{21}$ cm$^{-2}$ \citep{kalbet05}, so the measured absorption implies a large distance -- on the far side of the Galaxy or possibly in the halo \citep[see also][]{yaoet08}. Indeed, \cite{nowaet08} use \emph{Chandra}, \emph{XMM-Newton} and Rossi X-ray Timing Explorer (\emph{RXTE}) data to infer a rapidly spinning black hole and a distance of $\sim 10$--22 kpc. However, the authors do not explore the possibility of a neutron star using their data; instead they use the spectral and timing arguements above to favour a black hole.

\begin{table*}
\caption{Faulkes Telescopes observations of 4U 1957+11 used in this work.}
\begin{tabular}{llllllll}
\hline
Date      &MJD    &Camera  &Airmass&\multicolumn{3}{c}{Exposure times (sec)}\\
          &       &        &       &$V$	     &$R$      &$i$\\
\hline
2006-04-08&53833.6&DillCam &1.26   &120      &120      &120\\
2006-05-10&53865.6&DillCam &1.08   &200      &200      &200\\
2006-05-21&53876.6&DillCam &1.04   &200      &200      &200\\
2006-05-27&53882.6&DillCam &1.02   &200      &200      &200\\
2006-06-07&53893.6&DillCam &1.01   &200      &--       &--\\
2006-11-14&54053.2&HawkCam &1.24   &200      &--       &200\\
2007-04-22&54212.5&HawkCam1&1.56   &100      &--       &--\\
2007-06-16&54267.5&HawkCam1&1.09   &100      &100      &100\\
2007-07-10&54291.4&HawkCam1&1.20   &100      &100      &100\\
2007-07-12&54293.5&HawkCam1&1.01   &--       &--       &100\\
2007-08-06&54318.3&HawkCam1&1.12   &100      &100      &100\\
2007-08-10&54322.3&HawkCam1&1.09   &100      &100      &100\\
2007-08-19&54331.4&HawkCam1&1.03   &100      &100      &100\\
2007-09-02&54345.3&HawkCam1&1.02   &100      &100      &100\\
2007-10-05&54378.3&HawkCam1&1.08   &100      &100      &100\\
2007-11-13&54417.3&HawkCam1&2.42   &100      &100      &100\\
2008-06-13&54630.4&EM01    &1.70   &100      &100      &100\\
2008-06-16&54633.5&EM01    &1.02   & 80      &110, 100 &2 $\times$ 100 sec\\
2008-06-20&54637.6&EA01    &1.12   &100      &100      &100\\
2008-06-24&54641.5&EM01    &1.05   &--       &2 $\times$ 100 sec&2 $\times$ 100 sec\\
2008-06-25&54642.5&EM01    &1.01   &--       &100      &2 $\times$ 100 sec\\
2008-06-30&54647.6&EA02 (FTS)&1.20--1.42&--  &--       &18 $\times$ 100 sec\\
2008-07-11&54658.5&EM01    &1.01--1.04&2 $\times$ 100 sec&2 $\times$ 100 sec&16 $\times$ 100 sec\\
2008-07-14&54661.5&EM01    &1.04   &100      &100      &100\\
2008-07-14&54661.6&EA02 (FTS)&1.37   &2 $\times$ 100 sec&2 $\times$ 100 sec&4 $\times$ 100 sec\\
2008-07-16&54663.5&EM01    &1.19   &100      &100      &--\\
2008-07-16&54663.6&EA02 (FTS)&1.37--1.42&2 $\times$ 100 sec&100&3 $\times$ 100 sec\\
2008-07-30&54677.4&EA02 (FTS)&2.07   &200      &200      &200\\
2008-08-09&54687.3&EM01    &1.05   &100      &100      &100\\
2008-08-31&54709.5&EM01    &2.06   &100      &100      &100\\
2008-09-10&54719.5&EM01    &2.16--2.41&2 $\times$ 100 sec&100&2 $\times$ 100 sec\\
2008-09-15&54724.5&EM01    &2.13   &--       &100      &100\\
2008-09-25&54734.4&EM01    &1.52   &100      &100      &100\\
2008-11-04&54774.2&EM01    &1.18   &100      &100      &100\\
2009-02-22&54884.6&EM01    &2.50   &100      &100      &100\\
2009-05-17&54971.5&EM01    &1.12   &100      &100      &100\\
\hline
\end{tabular}
\small
\\ MJD = Modified Julian Day. The EA02 camera is on Faulkes Telescope South and all the other cameras are on Faulkes Telescope North. Data from 34 dates were used, totalling 144 images.
\normalsize
\end{table*}

An optical counterpart to 4U 1957+11 was discovered by \cite{marget78}, which was found to vary sinusoidally on an orbital period of 9.33 hours \citep{thor87}. However, a later photometric study revealed a more complex light curve of larger amplitude variability \citep*{hakaet99}. In this latter study the authors establish a colour-dependency in the amplitude; the optical (UBVRI) spectral energy distribution (SED) is slightly bluer at lower fluxes. \cite{hakaet99} modelled the light curve assuming it modulates on the orbital period, however the data cover just one 9-hour orbit. Since the optical behaviour is in contrast with the original sinusoidal modulation found \citep{thor87}, the variability they measure may not be periodic. \cite{shahet96} took a broad-band (4000--10,000\AA) optical spectrum of 4U 1957+11, which displayed a mostly featureless continuum, with weak H$\alpha$, H$\beta$ and He \small II \normalsize emission lines. No absorption lines from the companion star were detected. 

Here, we present the results of an optical monitoring campaign of 4U 1957+11 spanning 37 months during 2006--2009. Concurrent X-ray monitoring reveals a positive optical--X-ray correlation, and helps to constrain the origin of the optical variability and the nature of the compact object. The observations, reduction and flux calibration are described in Section 2. We analyze the results and give a comprehensive discussion, making comparisons to other sources, in Section 3. In Section 4 we summarize the results and conclusions.

\section{Observations}

\subsection{Optical photometry}

Observations of the field of 4U 1957+11 were taken with the Faulkes Telescope North situated at Haleakala on Maui, USA (FTN) and Faulkes Telescope South situated at Siding Spring in Australia (FTS). Both are 2-metre robotic telescopes optimized for research and education \citep*{robeet08,lewiro09}. Imaging was obtained in Bessell $V$, Bessell $R$ and Sloan Digital Sky Survey (SDSS) $i$
filters between April 2006 and May 2009, as part of a monitoring campaign of $\sim 30$ LMXBs \citep[see][]{lewiet08}. The log of observations is given in Table 1. Various cameras were used (column 3 of Table 1); all cameras have 2048 $\times$ 2048 pixels binned 2 $\times$ 2 prior to readout into effectively 1024 $\times$ 1024 pixels. For all cameras, the field of view is 4.7 $\times$ 4.7 arcmin and the pixel scale is 0.278 arcsec pixel$^{-1}$. Automatic pipelines de-bias and flat-field the Faulkes science images using calibration files from the beginning and the end of each night.

The seeing ranged from 0.7 to $> 3$ arcsec. We discarded images that were taken when the seeing was greater than six pixels (1.67 arcsec), because the measured magnitudes of field stars of similar magnitude to 4U 1957+11 became inaccurate under those poor seeing conditions. We also discarded images with poor signal-to-noise ratio (S/N; mostly due to thin cloud) and images with tracking problems (where the stars are non-spherical and stretch more than a few pixels during the exposure), an excess of hot pixels, or a background gradient across the image. 144 images were kept, which were taken between April 2006 and May 2009 (1138 days). The discarded images are not listed in Table 1.

We performed aperture photometry of 4U 1957+11 and two field stars using the package \small APPHOT \normalsize in \small IRAF \normalsize. Point-spread-function fitting was not used because minor tracking errors on some dates resulted in the stars being non-circular. A fixed aperture radius of 6 pixels was adopted (accurate centres to the stars are fitted), and a background annulus of 10--20 pixels was used.

Flux calibration was achieved using the standard star PG 1525--071 ($V$ = 15.053; $R$ = 15.141; $I$ = 15.218) from the list of Landolt photometric standards \citep{land92}, which is observed on FTN regularly in $V$, $R$ and $i$ filters. On 2006-04-08, 2007-03-22 and 2009-02-22 the conditions were photometric and both PG 1525--071 and 4U 1957+11 were observed by FTN. The SDSS $i$ magnitude of PG 1525--071 was calculated ($i$ = 15.53) adopting the prescription described in Table 3 of \cite*{jordet06} from the known $R$ and $I$ magnitudes. Taking into account the airmass differences between standard star image and target image on each of the three nights, we calculated the $V$, $R$ and $i$ magnitudes of two field stars in the target images. The stars are marked 1 and 2 in Fig. 1 and their magnitudes are listed in Table 2. Errors were estimated from the range in values from the three dates. Finally, to convert apparent magnitudes to intrinsic, de-reddened flux densities, we adopt A$_{\rm V}=0.93$ for the interstellar extinction towards the source \citep{marget78} and use the extinction curve of \cite*{cardet89}.

\begin{table}
\caption{Measured magnitudes of two nearby field stars (Fig. 1) used for flux calibration.}
\begin{tabular}{llll}
\hline
Star&$V$&$R$&$i$\\
1&16.47 $\pm$ 0.08&16.01 $\pm$ 0.08&15.85 $\pm$ 0.04\\
2&15.94 $\pm$ 0.08&15.35 $\pm$ 0.08&15.10 $\pm$ 0.03\\
\hline
\end{tabular}
\end{table}

\begin{table*}
\caption{Long and short-term flux and variability properties.}
\begin{tabular}{llllllll}
\hline
Waveband&Time span&Mean counts &1 $\sigma$&Mean relative&Mean unabs-&1 $\sigma$&Fractional\\
        &used     &or magnitude&range     &error        &orbed flux &range     &rms       \\
\hline
\multicolumn{8}{c}{\emph{4U 1957+11 long-term variability}}\\
1.5--12&13.5 &2.33 ASM cps&$\pm$ 1.03        &$\pm$ 0.42      &6.9$\times 10^{-10}$&$\pm$ 3.1$\times 10^{-10}$&44 $\pm$ 18 \%\\
keV    &years&            &ASM cps          &ASM cps        &ergs cm$^{-2}$ s$^{-1}$&ergs cm$^{-2}$ s$^{-1}$&\\
$V$    &3.1 years&18.93   &$\pm$ 0.23        &$\pm$ 0.04 	    &0.241 mJy	   &$\pm$ 0.050 mJy	    &21 $\pm$ 4 \%\\
$R$    &3.1 years&18.77   &$\pm$ 0.22        &$\pm$ 0.04 	    &0.170 mJy	   &$\pm$ 0.034 mJy	    &20 $\pm$ 3 \%\\
$i$    &3.1 years&18.88   &$\pm$ 0.21        &$\pm$ 0.04 	    &0.176 mJy	   &$\pm$ 0.033 mJy       &19 $\pm$ 4 \%\\
\hline
\multicolumn{8}{c}{\emph{Field star 3 long-term variability}}\\
$V$    &3.1 years&20.13   &$\pm$ 0.11        &$\pm$ 0.10 	    &\multicolumn{2}{c}{unknown extinction}&$< 19$ \%\\
$R$    &3.1 years&19.04   &$\pm$ 0.04        &$\pm$ 0.04 	    &\multicolumn{2}{c}{''}&$< 7$ \%\\
$i$    &3.1 years&18.42   &$\pm$ 0.04        &$\pm$ 0.03 	    &\multicolumn{2}{c}{''}&$< 7$ \%\\
\hline
\multicolumn{8}{c}{\emph{4U 1957+11 short-term variability}}\\
2--60 keV&2.0 years&\multicolumn{5}{c}{(From \citealt{wijnet02})}&1--2 \%\\
$i$    &0.9 hrs$^a$&19.18&$\pm$ 0.08&$\pm$ 0.07     &0.133 mJy	  &$\pm$ 0.010 mJy	   &7.3 $\pm$ 6.5 \% (1.1 $\sigma$)\\
$i$    &1.4 hrs$^b$&18.92&$\pm$ 0.05&$\pm$ 0.02     &0.169 mJy	  &$\pm$ 0.008 mJy	   &4.9 $\pm$ 2.2 \% (2.2 $\sigma$)\\
\hline
\multicolumn{8}{c}{\emph{Field star 3 short-term variability}}\\
$i$    &0.9 hrs$^a$&18.40&$\pm$ 0.05&$\pm$ 0.04     &\multicolumn{2}{c}{unknown extinction}&4.6 $\pm$ 3.4 \% (1.4 $\sigma$)\\
$i$    &1.4 hrs$^b$&18.41&$\pm$ 0.02&$\pm$ 0.02     &\multicolumn{2}{c}{''}&$< 3.0$  \%\\
\hline
\end{tabular}
\small
\\ The 2--60 keV data from \cite{wijnet02} are \emph{RXTE} Proportional Counter Array data from 1997-11-26 to 1999-12-02 but the variability timescales tested are 0.1--1000 seconds. All optical data are from the Faulkes Telescopes observations. $^a$data taken on 2008-06-30; $^b$data taken on 2008-07-11.
\normalsize
\end{table*}

\subsection{X-ray monitoring}

The All-Sky Monitor (ASM) on board the \emph{RXTE} satellite monitors the X-ray sky several times per day between 1.5 and 12 keV. 4U 1957+11 has been detected by the ASM since the start of its operation in 1996. The X-ray source has remained persistently active over the full 13.5 years: the mean daily count rate is 2.3 counts sec$^{-1}$ and the rms is 1.0 counts sec$^{-1}$. This range corresponds to an unabsorbed flux of (6.9 $\pm$ 3.1) $\times 10^{-10}$ ergs cm$^{-2}$ s$^{-1}$ (1.5--12 keV) assuming a blackbody at 1 keV (which is typical of an X-ray transient in the soft state) and an absorption of n$_{\rm H} = 1.5 \times 10^{21}$ cm$^{-2}$ \citep{nowaet08}. 86 per cent of ASM 1-day averages are 3 $\sigma$ detections, so it may be that on some dates the fluxes are significantly lower than the above range. We wish to cross-correlate optical and X-ray fluxes, so in order not to favour the brighter epochs, we did not discard the non-3 $\sigma$ detections. Instead, we binned the ASM data into 3-day averages, 5-day averages and so on to achieve higher S/N. This has the effect of smoothing the light curve; daily variability information is lost (for the analysis in Section 3 we study the strength of optical--X-ray correlations as a function of the number of ASM days used).

\begin{figure*}
\includegraphics[width=16cm,angle=0]{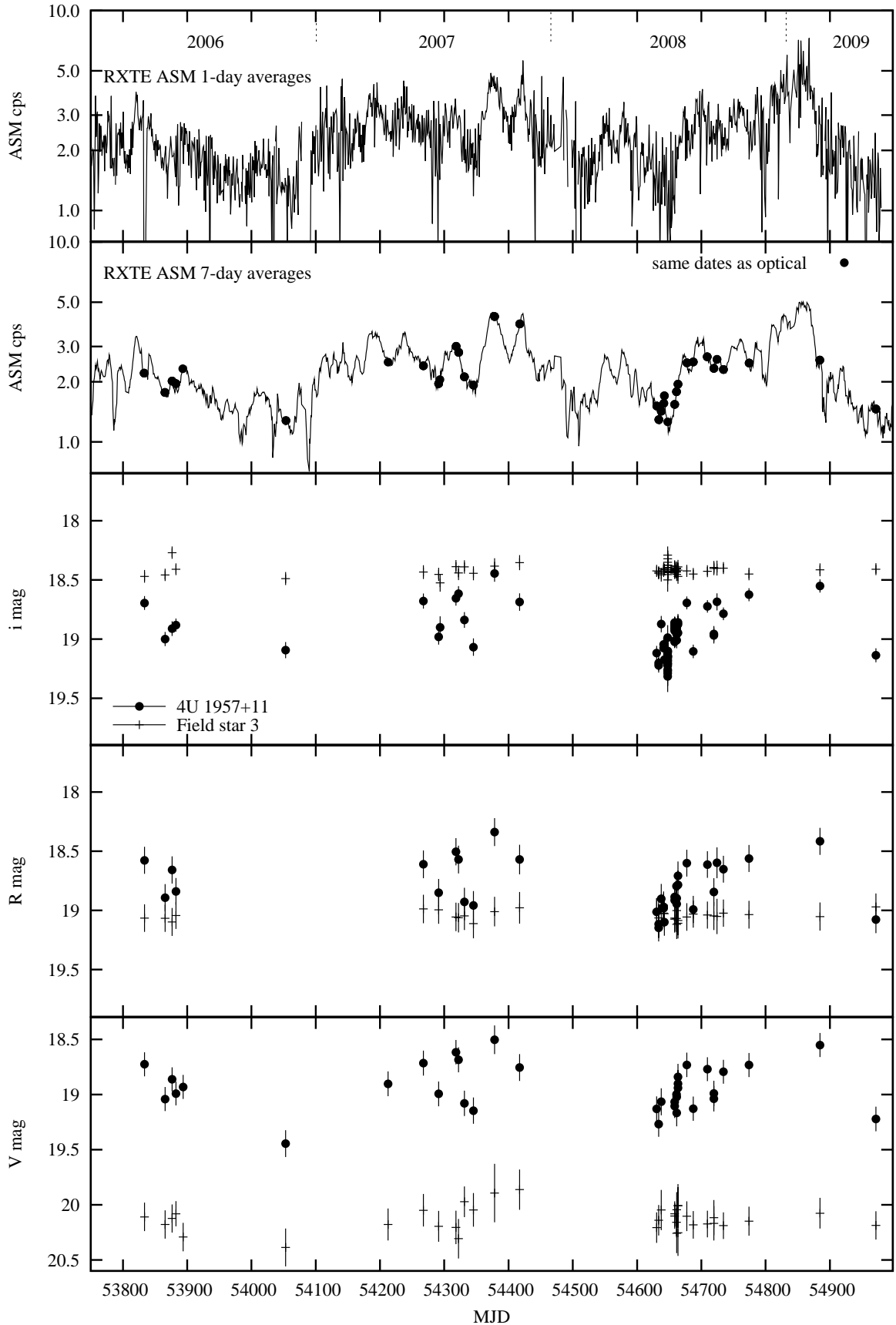}
\caption{X-ray (1.5--12 keV) and optical long-term light curves of 4U 1957+11. In the lower three panels, the magnitudes of 4U 1957+11 and a local field star are shown. The errors represent the total systematic and photometric magnitude errors.
}
\end{figure*}

\section{Analysis and discussion}

\subsection{Long-term variability}

The optical and X-ray light curves are presented in Fig. 2. It is clear from the ASM light curves (upper two panels) that the X-ray flux is dominated by long-term variations on tens to hundreds of days, but short-term variability on timescales of days or less is also present. Binning the ASM data into 7-day averages has the effect of smoothing the low-S/N 1-day averages light curve and improves the clarity of structure on tens-to-hundreds of days. In the lower three panels of Fig. 2, the $i$, $R$ and $V$ apparent (absorbed) magnitudes of 4U 1957+11 and a field star of similar magnitude are shown. The magnitude errors in the figure are the total errors; the relative errors are smaller. 4U 1957+11 is variable, spanning a range of almost one magnitude in all three filters.

From these light curves we derive the long-term X-ray and optical variability properties (Table 3). Over these long timescales (years) the fractional root mean square (rms) of the X-ray variability is $\sim 44$ per cent and in optical it is $\sim 20$ per cent (in all three filters). We may expect a smaller optical variability amplitude if the optical light arises from reprocessed X-ray photons on the surface of the accretion disc, as is often seen in LMXBs in outburst. The optical variability is real; the relative errors in the optical photometry ($\sim 0.04$ mag) are much smaller than the measured rms of the variability ($\sim 0.2$ mag). The long-term optical fractional rms of the field star is $< 7$ per cent; the errors in the magnitudes due to the photometry are similar to the apparent variability, so the field star is consistent with being non-variable. From Fig. 2 it appears the X-ray and optical fluxes could be correlated; we explore this possibility in Section 3.5.

\subsection{Short-term variability}

\cite{wijnet02} showed that the X-ray variability on 0.1--1000 second timescales is low, with amplitudes between one and two per cent. We tested for optical variability on timescales of minutes on two dates in 2008 (Fig. 3). We chose the $i$-band filter because the S/N is usually higher in $i$ than in $R$ and $V$ for a given exposure time. On 2008-06-30, 18 observations were made with FTS, with a time resolution of $\sim 170$ seconds. The source was faint ($i \sim 19.2$) and the S/N relatively low (likely due to a high airmass of up to 1.42). As a result, the relative magnitude errors were similar to or greater than the variability amplitude on this timescale, for both 4U 1957+11 and field star 3. On 2008-07-11 the source was brighter ($i \sim 18.9$), the conditions were favourable and the airmass very low. 15 observations were made with FTN within 36 minutes (the time resolution was $\sim 140$ seconds), and one 49 minutes before. It appears the source brightened during that time by $\sim 0.1$ mag. \cite{motcet85} also found 0.1 mag variations of the optical counterpart in one hour.

The properties of the optical and X-ray \citep[from][ten years previously]{wijnet02} short-term variability are also given in Table 3. The $i$-band variability on 2008-07-11 has a fractional rms of $\sim 3$--7 per cent. This is greater than the X-ray variability measured on these timescales, 1--2 per cent, however the $i$-band variability is significant at only the 2.2 $\sigma$ level. We do not detect variability from the field star, which is brighter than 4U 1957+11 in $i$-band.

\begin{figure}
\centering
\includegraphics[width=6cm,angle=270]{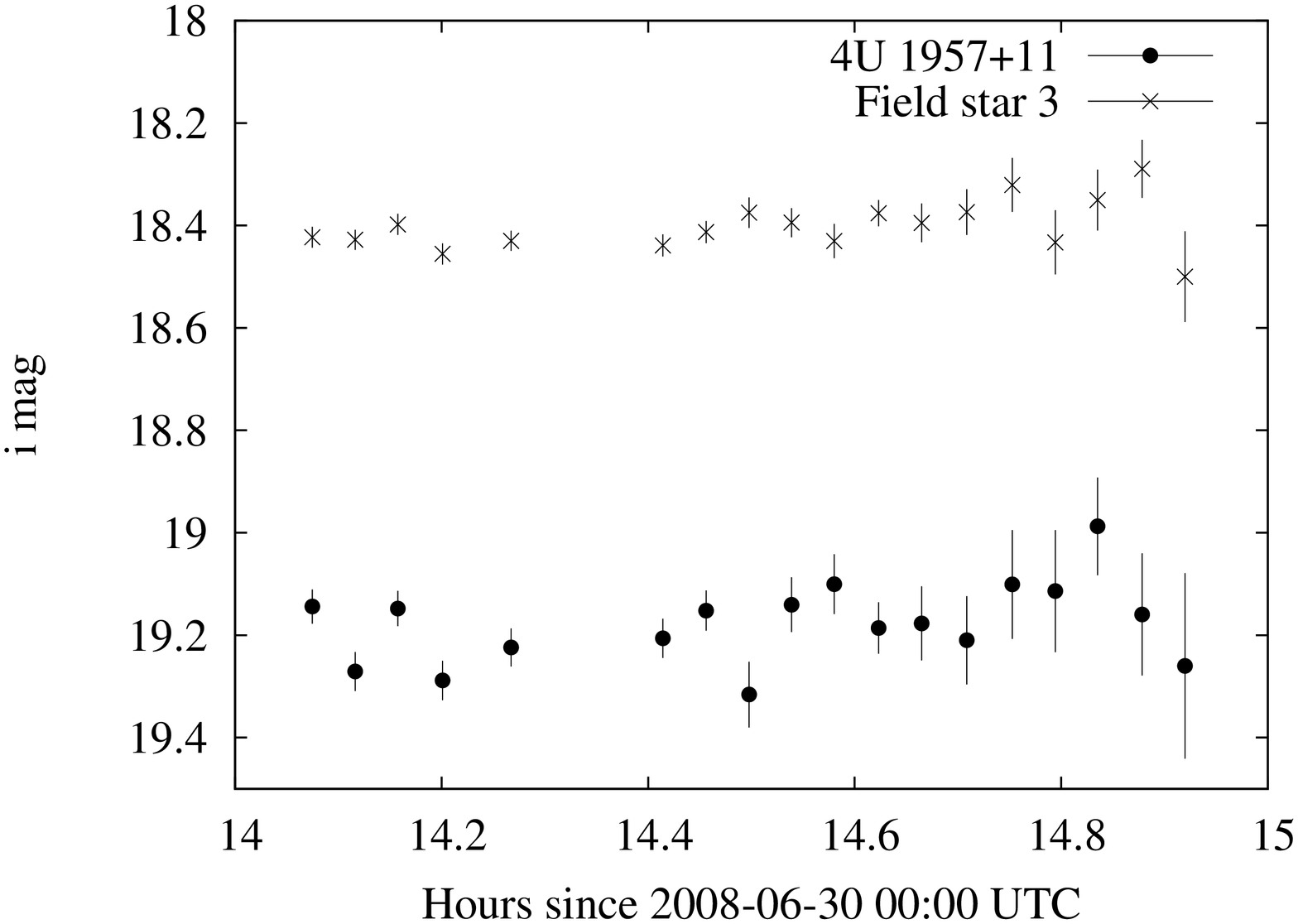}
\includegraphics[width=6cm,angle=270]{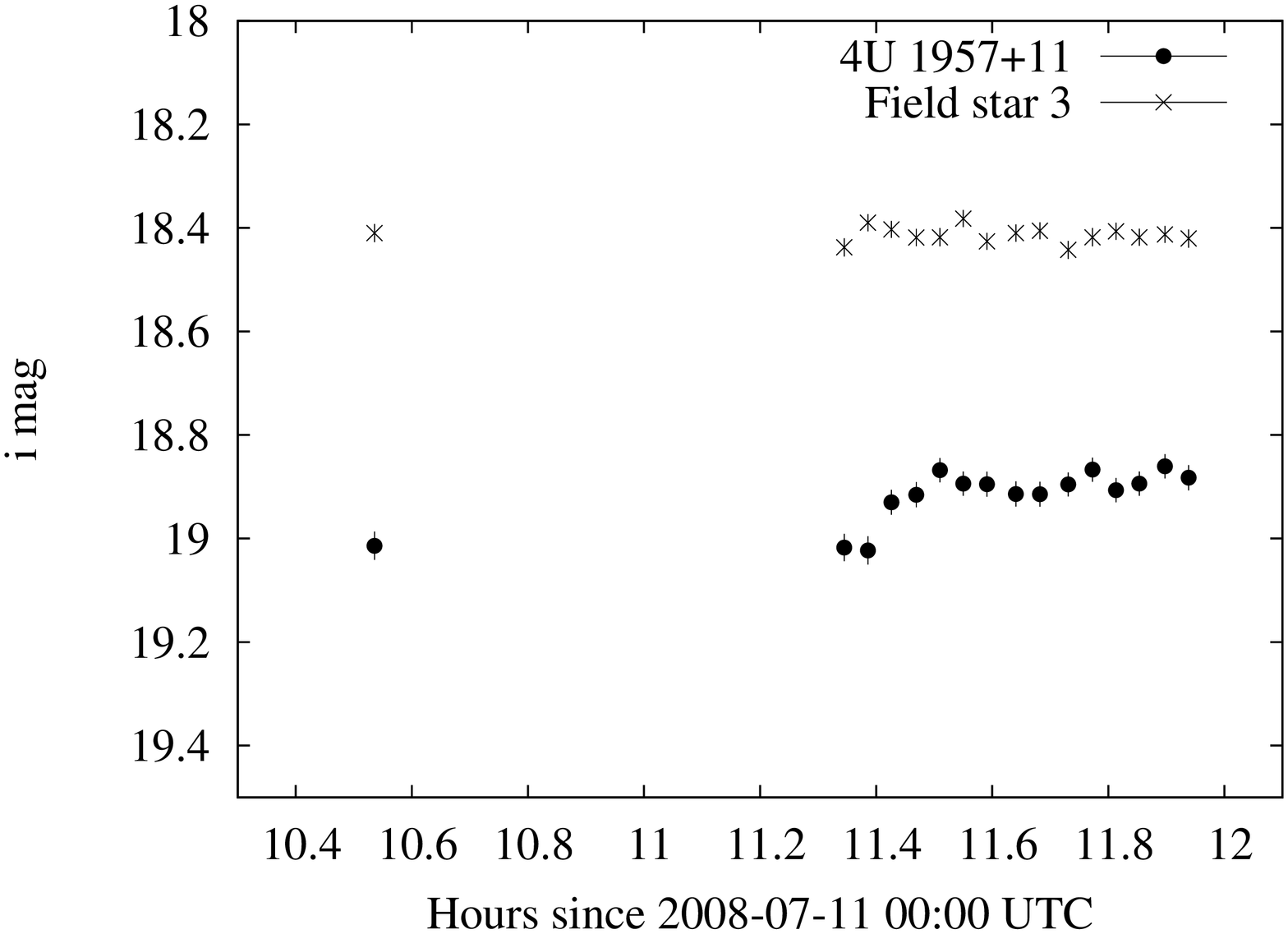}
\caption{$i$-band light curves with time resolutions of hundreds of seconds, on two dates in 2008.
}
\end{figure}

\begin{figure}
\centering
\includegraphics[width=6.17cm,angle=270]{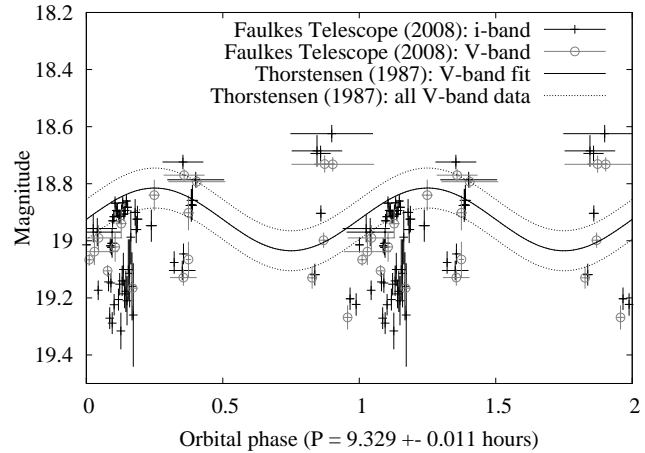}
\caption{Optical magnitude versus orbital phase (plotted twice for clarity), assuming the ephemeris of 
\citeauthor{thor87} (1987). Phase zero is arbitrary and does not correspond to a known orientation of the binary. The $V$ and $i$ magnitudes from June 2008 to September 2008 were used (outside this range the error of the orbital phase is too large).}
\end{figure}

\subsection{Periodic variability?}

We carried out a more intensive monitoring campaign of 4U 1957+11 in June -- September 2008, in order to search for periodic variability on the claimed orbital period of 9.329 $\pm$ 0.011 hours established in 1985 \citep{thor87}. We define zero phase at MJD 54646.0, at the centre of our campaign. This does not correspond to a known physical phase orientation of the binary. The above error of the period estimate propagates to an uncertainty of $\sim 100$ seconds per day, which results in an uncertainty in the relative phase which increases with distance in time from MJD 54646.0.

The folded light curve is presented in Fig. 4. We find no evidence for a modulation on this period, and the range of magnitudes in 2008 greatly exceeds that found 23 years previously by \cite{thor87}. Instead, the range of magnitudes we measure is consistent with those seen in 1996 by \cite{hakaet99}. Since \cite{thor87} discovered the periodicity from observations spanning 28 periods (11 nights), it is likely that the period was real at that time. The behaviour of the optical variability has clearly changed since 1985, as pointed out by \cite{hakaet99}. In the paper of \citeauthor{hakaet99} their observations covered just one 9.3-hour period, and it was assumed that the variations they witnessed were associated with this period. We suggest that the variability may not have been periodic at that time (see also Sections 3.5).

To test for the existence of other periodic signals, we performed timing analysis on our optical light curves using the \small STARLINK \normalsize package \small PERIOD\normalsize. No significant intrinsic signals were found in the $i$, $R$ or $V$ light curves using either the Lomb-Scargle \citep[][and references therein]{prery89} or epoch-folding \citep*[e.g.][]{hornet86} techniques, which are both appropriate for unevenly sampled data. For the $i$-band, in which we have short-term variability observations, we are able to search a frequency range of $\sim 10^{-8}$--$10^{-2}$ Hz (trial periods between $\sim 100$ seconds and 1 year). The strongest (but not highly significant) peaks are at periods of 24.4 hours, 49.8 days and 427.4 days. The Faulkes telescopes often observed the target around the same time each day during the monitoring, which probably causes the apparent 24.4-hour period. By folding the light curve on  49.8 days and 427.4 days, we see that these periods are dominated by the sampling of the two short-term variability studies, so they are also sampling aliasing artefacts. We find no evidence for optical periods of hundreds of days like those uncovered at X-ray energies \citep{nowawi99,wijnet02}. Follow-up dedicated monitoring campaigns covering various timescales would be beneficial for more stringent period searches.

\begin{figure*}
\centering
\includegraphics[width=8cm,angle=0]{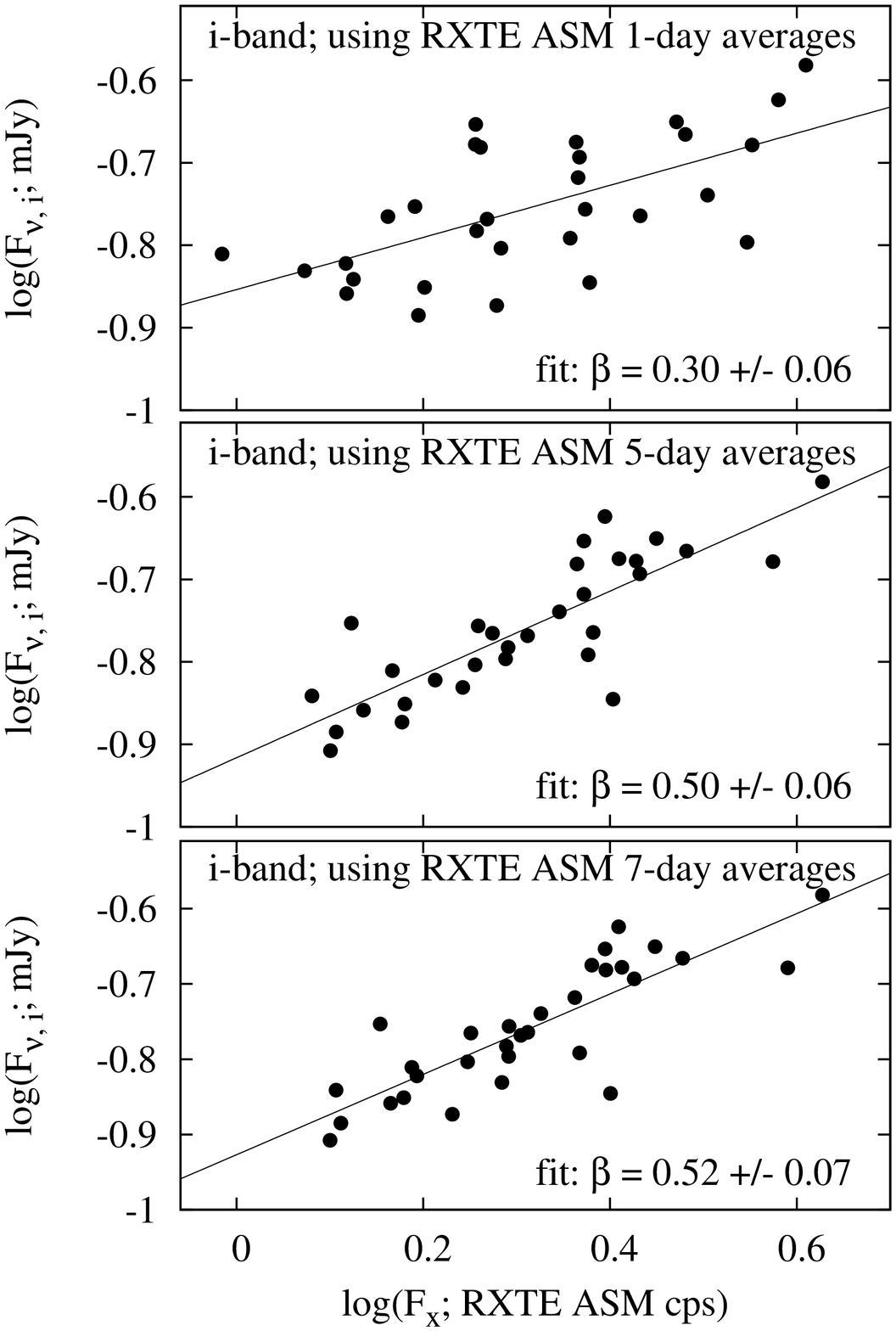}
\includegraphics[width=8cm,angle=0]{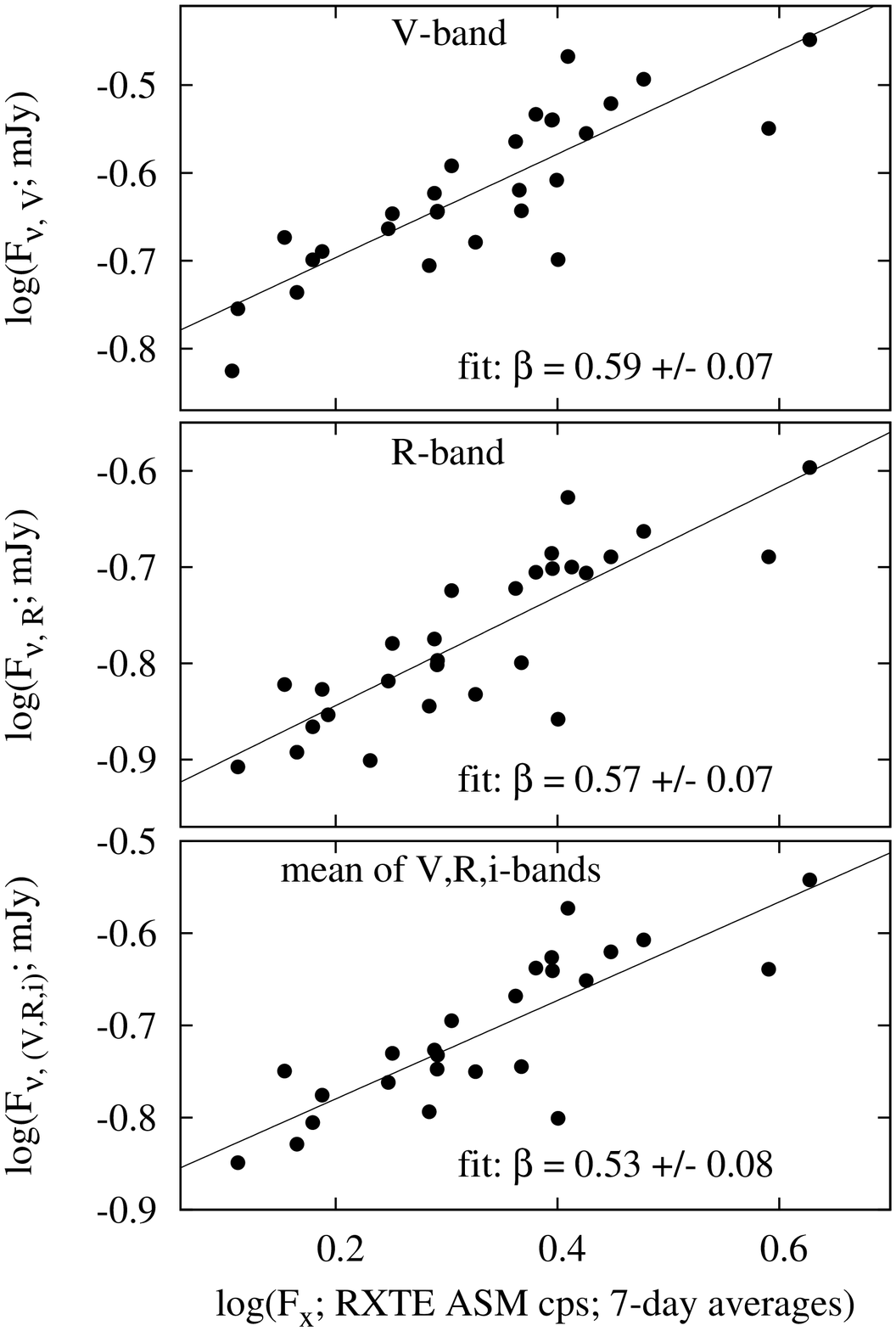}
\caption{Optical (de-reddened) versus X-ray fluxes centred on the same day. The solid lines are the best power law fits; the power law index $\beta$ is shown, where $F_{\rm \nu,OPT}\propto F_{\rm X}^{\beta}$. On the right panel we use \emph{RXTE} ASM 7-day means, which result in the strongest correlations (see text and Fig. 6).
}
\end{figure*}

\subsection{Optical spectral energy distribution}

The average optical ($V$ to $i$) spectral index over the three years of monitoring was $\alpha = +1.0$ (calculated from the mean de-reddened flux densities in Table 3), where $F_{\nu}\propto \nu^{\alpha}$. From all dates in which both $V$ and $i$ observations were made, the 1 $\sigma$ range of spectral indices observed was $\alpha = +0.92 \pm 0.25$. Since the extinction towards the source is known to be low (we adopt A$_{\rm V}=0.93$; see Section 2.1), our estimates of $\alpha$ are not likely to suffer large inaccuracies from an uncertain extinction. Systematic errors in the flux calibration could however produce errors in $\alpha$ up to 0.33.

This optical spectral energy distribution (SED) is typical of a LMXB in outburst \citep*{hyne05,russet07}. It is not consistent with the canonical $\alpha = \frac{1}{3}$ multi-temperature thermal disc spectrum, nor the Rayleigh-Jeans tail of a blackbody $\alpha = 2.0$ \citep*{franet02}. A redder SED would be expected if the emission were synchrotron in nature, as is seen in the optical/infrared (IR) SEDs of some LMXB jets \citep[e.g.][]{chatet03}. With $\alpha \sim +1$, we are likely sampling a region of the disc blackbody between the peak, which resides in the ultraviolet (UV), and the Rayleigh-Jeans tail in the IR \citep[e.g.][]{hyne05}. However, it is uncertain whether this is the Rayleigh-Jeans tail of the multi-temperature viscous disc (in which case $\alpha = \frac{1}{3}$ in the UV) or the irradiated disc \citep[where the disc surface reprocesses the X-ray photons to lower energies; here $\alpha = -1$ in the UV;][]{franet02}.

We would expect the optical SED to steepen (values of $\alpha$ to increase) with disc temperature, for a disc of constant area. This was recently shown to be the case for the optical--IR colours of a number of outbursts of Aql X--1 \citep{maitba08} and successfully modelled as a single-temperature heated blackbody. The $V$--$i$ colour of 4U 1957+11 is indeed bluer when it is brighter; in a forthcoming work we will apply the same technique to our data of 4U 1957+11 and other LMXBs \citep[Russell, Maitra et al. in preparation; for preliminary results see][]{russet08}.

\subsection{Optical--X-ray correlations}

If the accretion disc (viscously heated or irradiated) is producing the optical emission, as implied by the SEDs, we expect the optical and X-ray light curves to be correlated. Changes in the mass accretion rate (and possibly a precessing disc) produce the long-term X-ray variations \citep{nowawi99,wijnet02}. 
It has been shown \citep{vanpet94,russet06} that optical--X-ray correlations exist for LMXBs over orders of magnitude in luminosity. The power law index of the correlations depend on the process that produces the optical emission.

For 4U 1957+11, we take the mean magnitude from each day (0--24 UT) and plot the optical flux density against the X-ray 1.5--12 keV count rate centred on the same day, in Fig. 5. A positive optical--X-ray correlation exists which approximates a power law. The scatter in the correlation is smaller if we use 5 or 7-day ASM average count rates instead of 1-day averages (shown for $i$-band in the left panels of the figure). This is probably because the S/N of the 1-day averages is low, and the X-ray light curve is dominated by variations on timescales longer than one day. We do not include the errors associated with the optical and X-ray fluxes in fitting the correlations.

We investigate the strength of the correlation as a function of the number of days of ASM data averaged, in Fig. 6. The residuals to the fit (measured in log F$_{\rm \nu, OPT}$) are squared and summed, so the minimum in this plot corresponds to the least squares, and the strongest F$_{\rm OPT}$--F$_{\rm X}$ correlation. The best fit is obtained using 7-day ASM averages, for $V$ and $i$-bands, and for the mean of the $V$, $R$ and $i$-bands. The residuals are approximately twice as large if 1-day averages are used, confirming the low S/N of these data. If $> 9$-day averages are used, the residuals increase again. This is likely to be an indication of the timescale of the X-ray variability -- structure in the intrinsic X-ray light curve is diluted, or smoothed by averaging data over timescales longer than the variability. We therefore use 7-day ASM averages for the correlations presented in the right panel of Fig. 5.

To test the significance of the correlations we make use of the Spearman's Rank correlation. A correlation exists between F$_{\rm OPT}$ and F$_{\rm X}$ at the 4.4 $\sigma$, 4.3 $\sigma$, 4.5 $\sigma$ and 4.0 $\sigma$ levels for $V$, $R$, $i$ and the mean of $V$,$R$,$i$-bands, respectively. The significance of the correlation is highest when $i$-band is used, for which we have the most data points, and least when the mean of the three filters is used, for which we have the least number of data points. Since the variability timescale is on the order of days, these significances may be overestimated by many data points separated by less than this timescale. We therefore calculate the Spearman's Rank correlation again for $i$-band, removing all data within 20 days of an other data point. Data on 18 dates were left, for which a Spearman's Rank correlation exists with a significance of 3.1 $\sigma$.

The fits to the correlations have power law indices of $\beta = 0.5$--0.6 (where F$_{\rm OPT} \propto$ F$_{\rm X}^{\beta}$), which is empirically typical for both BHXBs and NSXBs over several orders of magnitude \citep{russet06,russet07}. Within errors, $\beta$ is no different if $V$, $R$, $i$-band data or the mean of the three are used. $\beta \approx 0.5$ is expected if X-ray reprocessing on the surface of the disc produces the optical emission (\citealt{vanpet94}; although see \citealt*{dubuet01} for a relation with an evolving $\beta$ based on modelling light curve morphologies). The same relation is also expected if the viscously heated disc dominates, but only in the case of a NSXB; $\beta$ is expected to be shallower for a BHXB \citep{russet06}. Optical synchrotron emission from a BHXB jet would produce a slightly steeper power law, $\beta \approx 0.7$ \citep{russet06}; although the measured power law index for 4U 1957+11 is close to this value, the spectral index is inconsistent with a synchrotron origin (Section 3.4).

\begin{figure}
\centering
\includegraphics[width=6cm,angle=270]{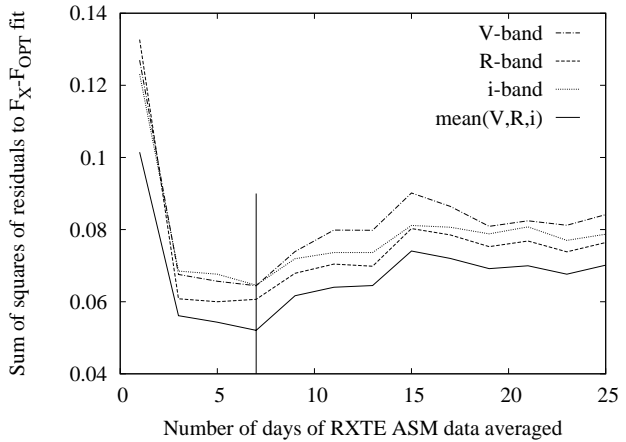}
\caption{The strength (least squares) of the $F_{\rm OPT}$--$F_{\rm X}$ correlation versus the number of days of averaged \emph{RXTE} ASM data used (centred on the date of each optical observation). The vertical line indicates seven days of averaged X-ray data, for which the strongest $F_{\rm OPT}$--$F_{\rm X}$ correlations exist.
}
\end{figure}

So far we have investigated optical--X-ray correlations in which the X-ray data are centred on the same day as the optical. It is also possible to test whether one of the light curves lags the other. The discrete cross-correlation function \citep[CCF;][]{edelkr88} is the standard, effective method for two unevenly sampled light curves. In Fig. 7 we plot the discrete CCF for optical and X-ray fluxes, using $i$-band (upper panel) and the mean of $V$, $R$ and $i$-bands (centre panel). We take the mean optical magnitude per filter on each day in which two or more observations were made.

\begin{figure}
\centering
\includegraphics[width=6cm,angle=270]{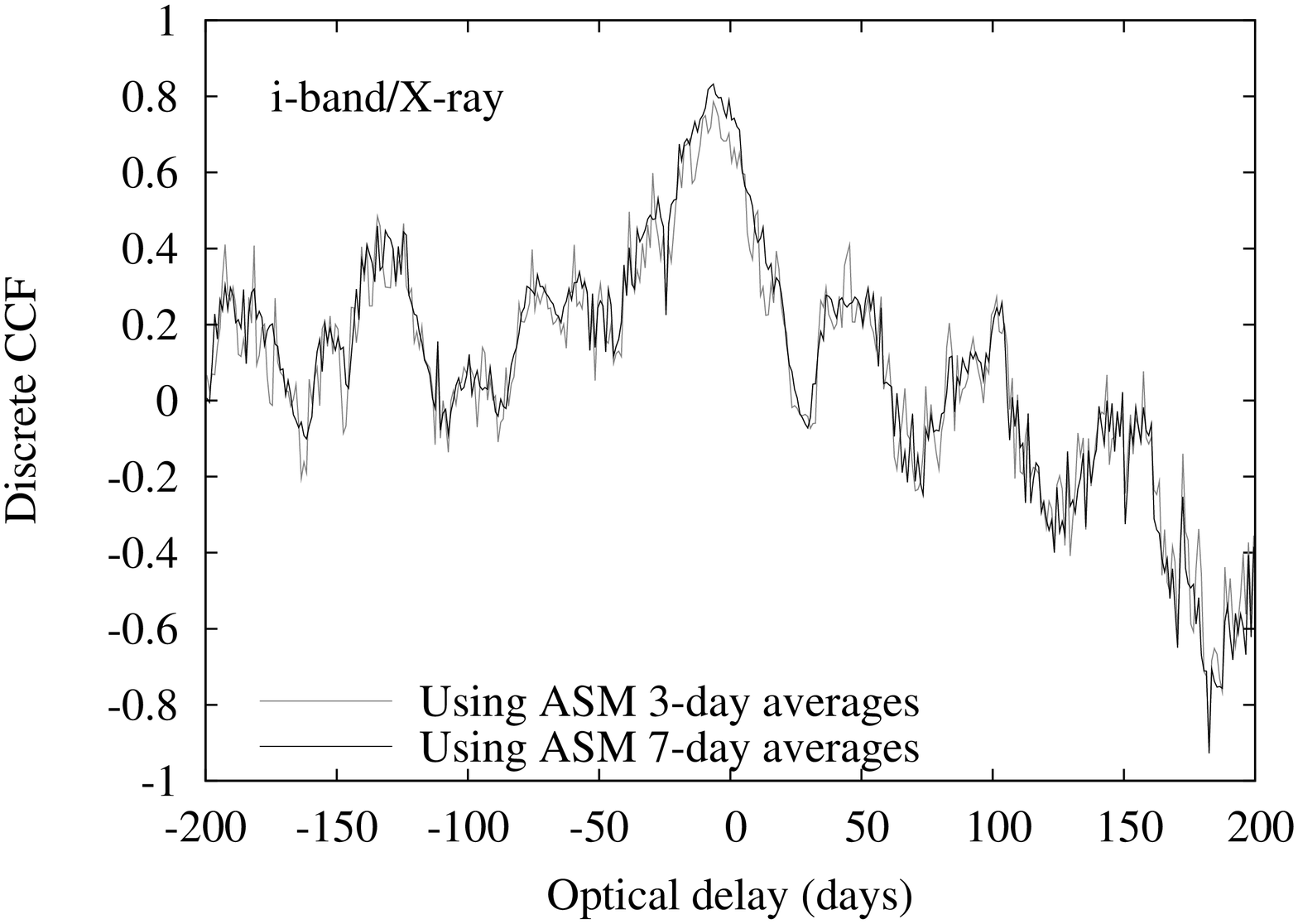}
\includegraphics[width=6cm,angle=270]{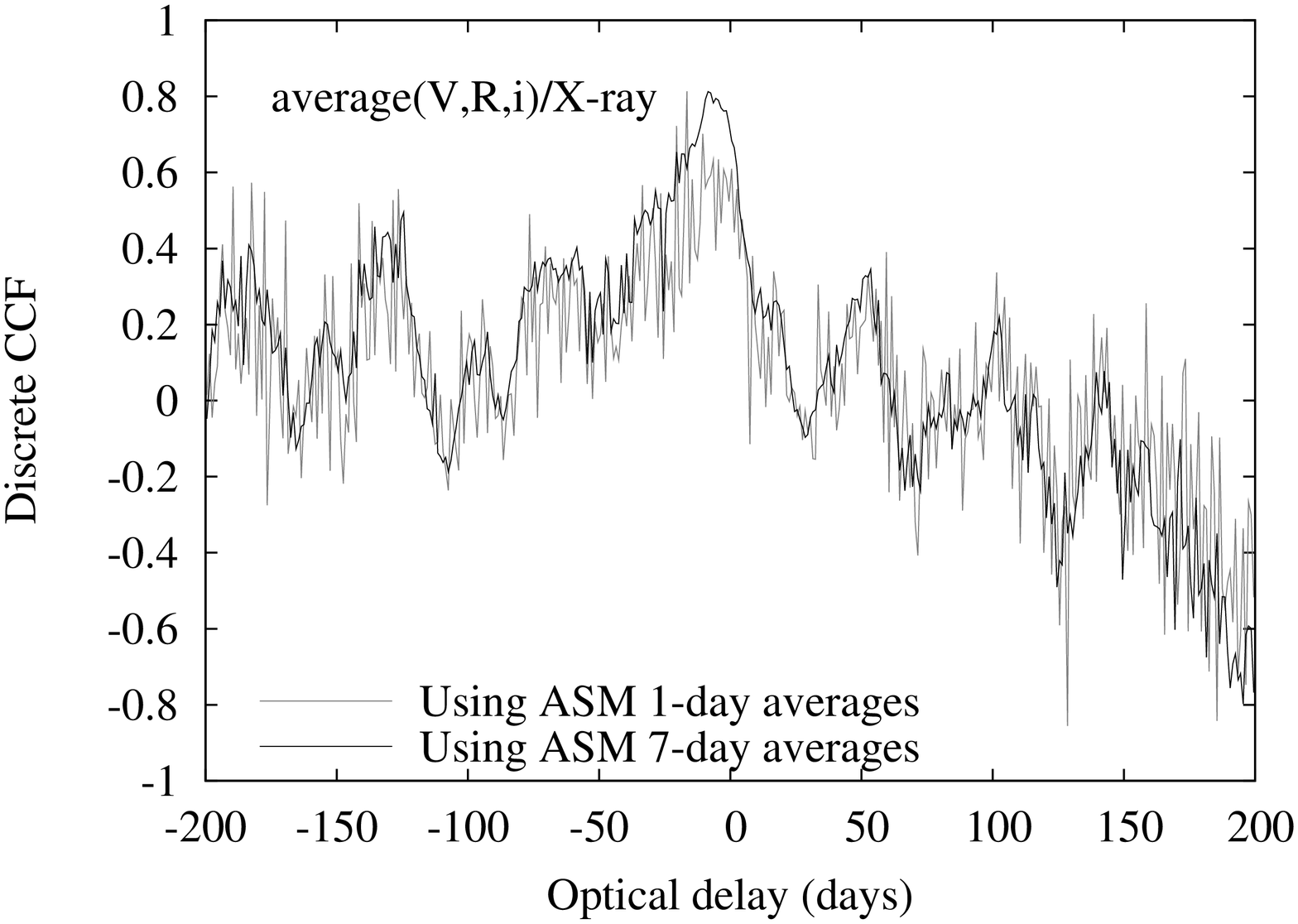}
\includegraphics[width=6cm,angle=270]{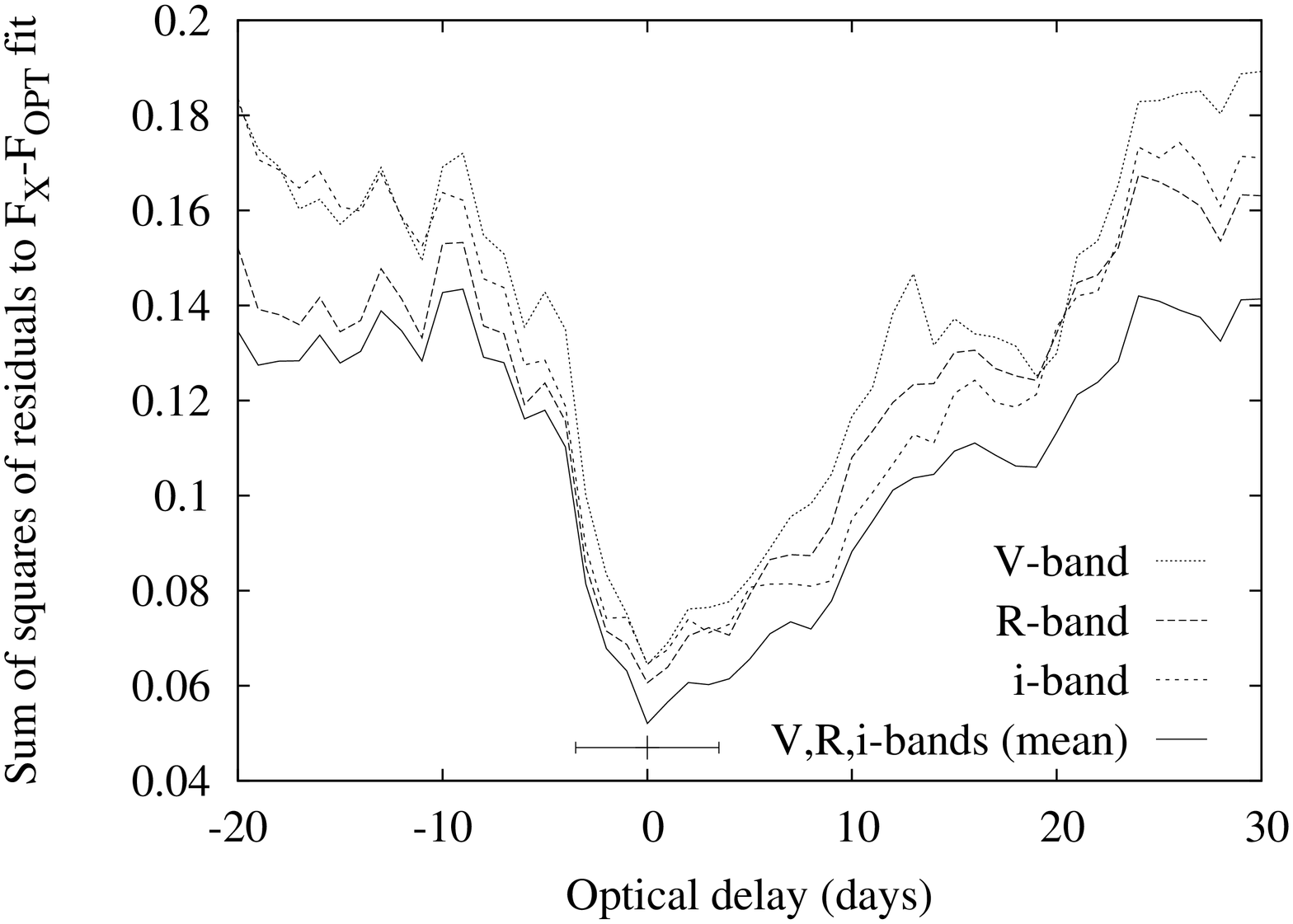}
\caption{The discrete CCF using $i$-band/1.5--12 keV (upper panel) and mean($V$,$R$,$i$)/1.5--12 keV (centre panel). Lower panel: the strength of the fit to the optical--X-ray correlation, as measured from the sum of the squares of the residuals of the fit, versus the optical delay (see text). The error bar at the bottom of the plot represents the 7-day range of X-ray data used.
}
\end{figure}

A positive correlation is indeed seen, peaking at an X-ray lag of $\sim 7$ days. We estimated the error associated with this lag thus. We removed one third of the $i$-band data (10 random data points from the 32) and performed the discrete CCF again, measuring the peak lag. 25 iterations of this were carried out using $i$-band and ASM 7-day averages (first trial) and 25 further iterations using $i$-band and ASM 3-day averages (second trial). The mean lag and 1 $\sigma$ error were $7 \pm 1$ days for the first trial and $8 \pm 4$ days for the second trial. In none of the 25 iterations was an optical lag measured in either trial (the X-ray lag was always greater than zero). Since seven days of averaged ASM fluxes were used for each optical data point in the first trial, the formal error on the measured lag in that trial cannot be less than $\pm 4$ days. Likewise for the second trial the error on the lag must be no less than $\pm 2$ days. We therefore measure X-ray lags of $7 \pm 5$ days from the first trial and $8 \pm 6$ from the second trial. Using the most conservative of these errors, we derive a measured X-ray lag of 2--14 days from the discrete CCFs.

For X-ray reprocessing, we would expect an optical lag on light-travel timescales \citep[seconds; see e.g.][]{obriet02}, i.e. consistent with zero days lag. The optical precedes the X-ray by several days, which is indicative of optical variations from the intrinsic viscous disc. X-ray delays of several days are expected if matter or waves of turbulence propagate inwards from the optically-emitting outer disc to the X-ray-emitting inner disc. This implies that for 4U 1957+11, accretion rate perturbations in the disc take 2--14 days to propagate from the optically-emitting outer disc to the X-ray-emitting inner regions close to the compact object.

\cite*{brocet01} find a similar discrete CCF for the BHXB LMC X--3, with an X-ray lag of 5--10 days from long-term optical and X-ray light curves. 4U 1957+11 and LMC X--3 share many properties -- both are persistent with a predominantly soft X-ray spectrum. It was found that both the optical and X-ray long-term variations of LMC X--3 are driven by changes in the mass accretion rate \citep{wilmet01,brocet01}. We arrive at the same conclusion for 4U 1957+11. It is also worth noting that the X-ray light curve of the NSXB 4U 0614+09 also varies with the mass accretion rate \citep{vanset00} and the optical variability may be correlated with the X-ray \citep{machet90,shahet08}. In addition, \cite{homaet05} found an X-ray lag of 15--20 days compared to near-IR variations in the BHXB GX 339--4 during a soft state. These measured lags probe the disc viscosity; they may be a function of both the size and viscosity of the disc.

A further test of optical or X-ray lags is to directly calculate the strength of the optical--X-ray correlation as a function of the lag. We fit the F$_{\rm OPT}$--F$_{\rm X}$ correlation taking X-ray data (7-day averages) centred at a series of days before and after each optical observation. We found that the correlation was strongest (the sum of the squares of the residuals was least) when X-ray data centred on the same day as the optical were used (independently for all three optical filters). The error associated with this lag is $\pm 4$ days because ASM 7-day averages were used. The strength of the F$_{\rm OPT}$--F$_{\rm X}$ correlation is plotted against the optical delay in the lower panel of Fig. 7. The correlation is not strong for an X-ray lag of $> 4$ days. The discrete CCF peaked at an X-ray lag of 2--14 days; this lag is not reproduced here using the least squares method.

The main difference between the two methods (discrete CCF and least squares to the F$_{\rm OPT}$--F$_{\rm X}$ correlation) is that the discrete CCF is normalised by the variance of the light curves (the product of the standard deviations). If the variability characteristics are very different on different timescales (as would be expected if there were two distinct processes contributing to the variability), the shape of the resulting correlation function will be more sensitive to the correlation that is strongest on that particular timescale. If this is the case then the discrete CCF is sensitive to the variability on timescales of days (such as the viscous timescale) and the least squares method is most sensitive to timescales no greater than $\sim $ one day (light travel timescales are $\ll$ one day). In addition, the least squares method tests for a power law relation between F$_{\rm OPT}$ and F$_{\rm X}$. This method would therefore be ineffective if the two fluxes are correlated but not by a single power law.

\begin{figure*}
\centering
\includegraphics[width=12cm,angle=270]{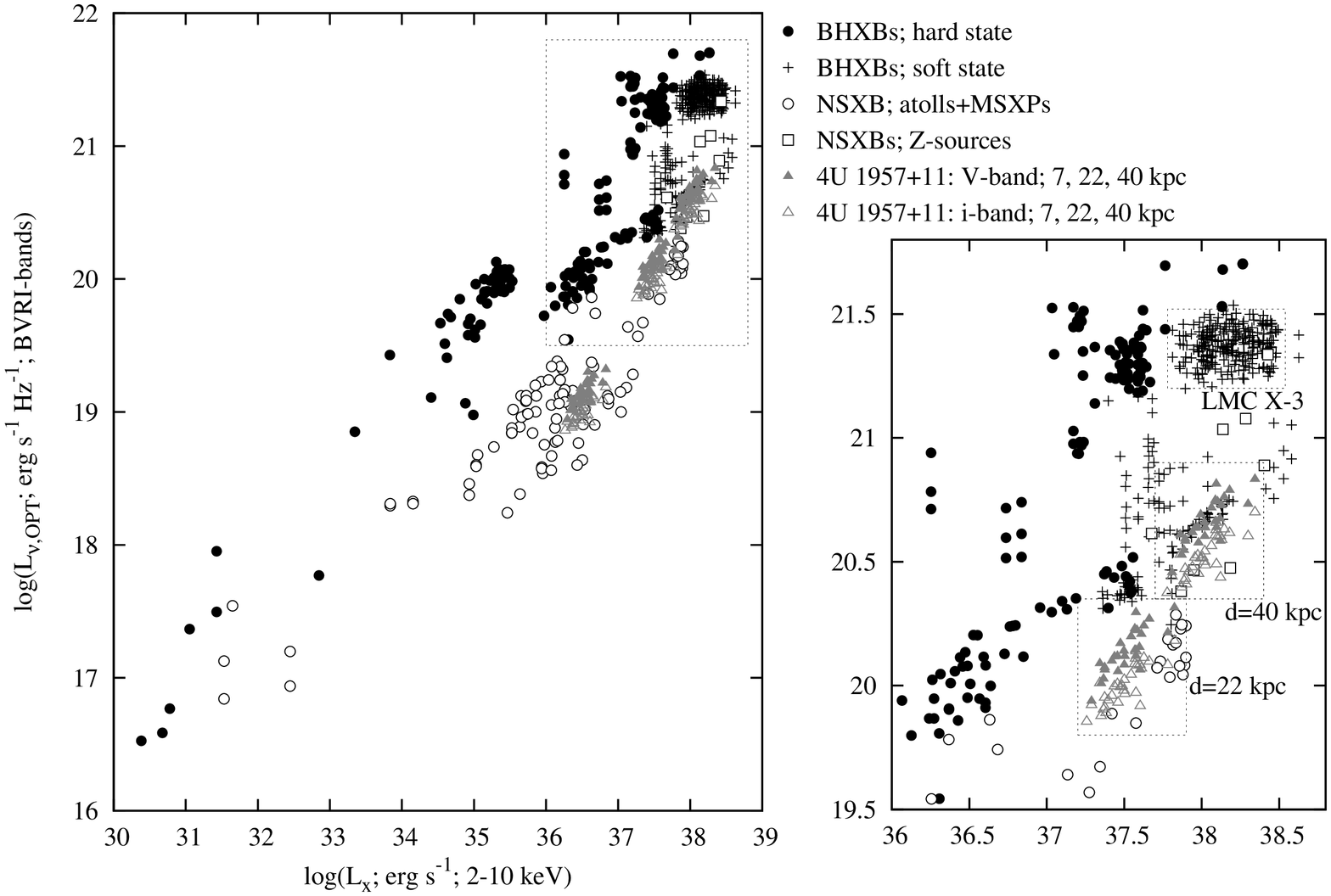}
\caption{X-ray luminosity versus monochromatic optical luminosity for all BHXBs and NSXBs collected \citep{russet06,russet07}. The neutron star `Z-sources' are plotted as separate symbols to neutron star `atoll' sources \citep[for definitions see][]{hasiva89} and millisecond X-ray pulsars. The $V$- and $i$-band data for 4U 1957+11 are overplotted assuming three distances encompassing the minimum and maximum likely values. In the left panel all luminosities are plotted, from quiescence to peak outburst; the high luminosity region is expanded in the right panel.
}
\end{figure*}

Alternatively, the two methods may simply propagate errors systematically in different ways, resulting in different measurements for the same lag. The two methods result in lags that indeed overlap (both are consistent with an X-ray lag of 2--4 days) and since the systematic errors using each method are unknown, we must take into account all the errors (especially since the S/N of the X-ray data is low). We therefore conclude that the optical lag is constrained to be between $-14$ and $+4$ days.

Signatures of both phenomena -- variability from the viscous disc and the X-ray heated disc -- are also found in the optical emission of active galactic nuclei \citep[AGN; see][and references therein]{breeet09}. Here, the larger size scales of the accretion disc and supermassive black hole result in reprocessing timescales of days instead of seconds and viscous timescales of months--years instead of days. Like LMXBs, X-ray reprocessing tends to be the dominating source of optical variability in AGN, but years of monitoring of some AGN have uncovered long-timescale variations from mass accretion rate changes or thermal fluctuations in the disc \citep*[e.g.][]{shemet03,marset08,arevet09}. One difference appears to be that the amplitude of the very long-term optical variations in AGN exceeds that of the X-ray, whereas in LMXBs (at least for 4U 1957+11 and LMC X--3) the long-term X-ray variations are of higher amplitude than the optical.

\subsection{A black hole or a neutron star system?}

The nature of the compact object in the 4U 1957+11 system is disputed (Section 1). \cite{russet06} showed from data collected from 24 LMXBs that BHXBs and NSXBs reside in different regions of the optical--X-ray luminosity diagram (Figures 1 and 2 in their paper). It would therefore be fruitful to utilize this as a diagnostic diagram to constrain the BHXB/NSXB nature of 4U 1957+11. In Fig. 8 we plot the quasi-simultaneous X-ray luminosity and optical (BVRI-bands; we exclude all near-IR data) monochromatic luminosity (flux density scaled to distance) for all BHXBs and NSXBs collected in \cite{russet06,russet07}. The distance to 4U 1957+11 is unknown, so we overplot the optical--X-ray data of 4U 1957+11 adopting three trial distances. The minimum distance is $\sim 7$ kpc and from high-resolution X-ray spectra the source is consistent with a BHXB at 10--22 kpc \citep{marget78,nowaet08}. If the source is in the Galactic halo its distance could be $\sim 40$ kpc \citep{yaoet08}.

At a distance of 7 kpc, the source lies amongst the NSXBs in the diagram and has an X-ray luminosity of $10^{36-37}$ erg s$^{-1}$ (2--10 keV). All BHXBs are $\sim$ one order of magnitude brighter in optical at this X-ray luminosity (and are hard state BHXBs not soft state BHXBs), so this would strongly favour a NSXB. At 22 kpc, the data still lie in a region of the diagram populated by the NSXBs, but almost overlap with some of the soft state BHXB data (we should not compare with the hard state BHXB data because 4U 1957+11 is a soft state source; see the right panel of Fig. 8 for an expanded view of this region of the diagram). If 4U 1957+11 is in the halo at 40 kpc, the data are typical of a (fairly optically faint) soft state BHXB, or a `Z-source' NSXB. At 40 kpc the source may be super Eddington; for a 10 M$_{\odot}$ black hole the bolometric luminosity observed would be twice the Eddington luminosity. We note that LMC X--3 is optically brighter than 4U 1957+11 (Fig. 8; right panel).

We propose three scenarios for the nature of 4U 1957+11:

\begin{itemize}
\item It is a BHXB on the far side of the Galaxy at d$\approx 20$--25 kpc. This would be consistent with both the X-ray timing/spectral evidence for a BHXB \cite[e.g.][]{wijnet02} and (marginally) the optical/X-ray luminosities of a soft state BHXB (Fig. 8);
\item It is a halo BHXB at $d > 25$ kpc. The optical/X-ray luminosities would be typical of other soft state BHXBs in this case but BHXBs in the halo are unusual \citep{whitva96}. This scenario would have important implications for models of supernova `kicks' and BHXB formation \citep[e.g.][]{miraet01};
\item The source is in the Galaxy, at d$\approx 7$--20 kpc. It would then be peculiar: it would either be an anomalous NSXB with X-ray properties like no other NSXB, or the faintest soft state BHXB, with the lowest normalization in the optical--X-ray luminosity diagram.
\end{itemize}

\section{Conclusions}

We have monitored the optical flux of the persistent LMXB 4U 1957+11 in three wavebands over a period of three years with the two 2-m Faulkes Telescopes. Like the soft X-ray light curve, the optical continuum is dominated by long-term variations on timescales of days--months, likely originating in changes of the mass accretion rate. The amplitude is a factor $\sim$ two in flux over these timescales; around half the amplitude of the X-ray variations. Contrary to previous shorter-timescale studies, no clear periodicities exist in the optical light curves. However we do not dispute the orbital period of 9.3 hours previously identified over a time span of $\sim 10$ days by \cite{thor87}. We suggest the X-ray flux (and so X-ray heating of the disc and star) during their observations may have been fairly stable (as is sometimes the case on those timescales; see Fig. 2) so that the low-amplitude sinusoidal variations from the heated face of the companion star were revealed (to our knowledge no X-ray data were acquired during that epoch). Under the assumption that the variability they see over one 9-hour period in 1996 was periodic, \cite{hakaet99} model their complex light curve as an evolving accretion disc, and infer a mass ratio $q \approx 0.4$--0.5 and an inclination angle of 70--75$^{\circ}$. We remark that these estimates of the system parameters are invalid as the flux was probably not periodic at that time.

The optical SED is blue ($\alpha \approx +1.0$) and slightly redder at lower fluxes. This is indicative of an accretion disc (heated viscously or by irradiation) and rules out synchrotron emission (which has an even redder SED) as the dominating process. We establish a positive correlation between the optical and \emph{RXTE} ASM 1.5--12 keV X-ray fluxes, of $F_{\rm \nu,OPT}\propto F_{\rm X}^{0.5}$, significant at up to the 4.5 $\sigma$ level (conservatively, $> 3 \sigma$) and evident from the three optical wavebands independently. The power law index of the correlation shows that the empirical correlation found over whole outbursts for LMXBs \citep{russet06} can hold over much smaller ranges in luminosity.

Cross-correlating the optical and X-ray light curves with the least squares method reveals a lag of $0 \pm 4$ days, consistent with X-ray reprocessing in which X-ray variations should precede the optical on light-travel timescales (seconds). However, the peak in the discrete CCF is at an X-ray lag of 2--14 days. This could be the viscous timescale for matter or waves of turbulence to travel from the optically-emitting outer disc to the X-ray-emitting inner disc. Combining these two methods we constrain the optical lag to be between -14 and +4 days. Optical and X-ray monitoring with more data and with higher S/N in X-ray is required to further constrain these lags.

BHXBs and NSXBs lie in different regions of the optical--X-ray luminosity diagram. We therefore overplot the data of 4U 1957+11, adopting a range of distances to the source, in order to constrain the currently disputed nature of the compact object. At distances $< 20$ kpc the data lie in a region of the diagram populated only by neutron star sources. If it were a BHXB at this distance, the optical emission would be one order of magnitude fainter than the other BHXBs, and it would be the faintest soft state BHXB. We therefore postulate that if its distance is $< 20$ kpc it is most likely a peculiar neutron star system.

However, taking into account the BHXB-like X-ray properties of 4U 1957+11, the most likely scenario is that the source is a distant BHXB. Its optical and X-ray luminosities are consistent with other soft state black hole systems if its distance exceeds $\sim 20$ kpc. 4U 1957+11 may well be a Galactic halo BHXB, in analogy with the prototypical halo BHXB, XTE J1118+480. This would imply a high space velocity and a large supernova `kick' at its birth. Optical phase-resolved radial velocity studies during a faint period (when the star/disc ratio is highest) may be required to identify the compact object unambiguously.

If 4U 1957+11 indeed contains a black hole, it is a valuable addition to the very few persistently accreting black hole X-ray binaries known (e.g. LMC X--1; LMC X--3) and possibly only the second in our Galaxy, after Cygnus X--1. In addition, 4U 1957+11 would be the perfect candidate to test for the presence of a weak soft state radio jet. A steady, core radio jet has never been detected from a BHXB in the soft state \citep[e.g.][]{fendet09} but it is speculated that faint jets (either intrinsically weak or with very low radiative efficiency) may be produced by soft state BHXBs. The only radio observations of 4U 1957+11 in the literature provide an upper limit of 2.1 mJy at 5 GHz \citep{nelssp88}. X-ray binaries can be detected down to, for example 50 $\mu$Jy or fainter with the Very Large Array \citep{gallet06} and much fainter with future planned Square Kilometer Array Pathfinders.

\section*{Acknowledgments}

We thank Phil Uttley, Monihar Dillon and Tom Maccarone for insightful discussions. The Faulkes Telescope Project is an educational and research arm of the Las Cumbres Observatory Global Telescope (LCOGT). DMR acknowledges support from a Netherlands Organization for Scientific Research (NWO) Veni Fellowship. FL acknowledges support from the Dill Faulkes Educational Trust.

\bsp

\label{lastpage}

\end{document}